\documentclass{article}
\usepackage{graphicx}

\begin{document}

\begin{center}
\Large {\bf Magnetic state in  URu$_{2}$Si$_{2}$,
UPd$_{2}$Al$_{3}$ and UNi$_{2}$Al$_{3}$ probed  by point contacts}

\vspace {1cm}

\large
 Yu. G. Naidyuk$^{a,b}$, O. E. Kvitnitskaya$^{a,b}$, A. G.
M. Jansen$^{a}$, C. Geibel$^{c}$, A. A. Menovsky$^{d}$, P.
Wyder$^{a}$

\end{center}
\vspace {1cm}

 {\it $^{a}$Grenoble High Magnetic Field Laboratory,
Max-Planck-Institut f\"ur Festk\"or\-perforschung and Centre
National de la Recherche Scientifique, Grenoble Cedex 9, F-38042,
France\\

$^{b}$B. Verkin Institute for Low Temperature Physics
and Engineering, NAS of Ukraine, 61164 Kharkiv, Ukraine\\

$^{c}$Max-Planck Institut f\"ur chemische Physik fester Stoffe,
Dresden, D-01187, Germany\\

$^{d}$ Van der Waals-Zeeman Laboratory, University of Amsterdam
1018 XE, The Netherlands}

\vspace {1cm}
\begin{abstract}
The antiferromagnetic (AFM) state has been investigated in the
three heavy-fermion compounds URu$_2$Si$_2$, UPd$_2$Al$_3$, and
UNi$_2$Al$_3$ by measuring d$V$/d$I(V)$ curves of point contacts
at different temperatures (1.5-20\,K) and magnetic fields
(0-28\,T). The zero-bias maximum in d$V$/d$I(V)$ for
URu$_{2}$Si$_{2}$ points to a partially gapped Fermi-surface
related to the itinerant nature of the AFM state contrary to
UPd$_{2}$Al$_{3}$ where analogous features have not been found.
The AFM state in UNi$_{2}$Al$_{3}$ has more similarities with
URu$_{2}$Si$_{2}$. For URu$_{2}$Si$_{2}$, the same  critical
field of about 40~T along  the  easy c axis is found for all
features in d$V$/d$I(V)$ corresponding to the N\'eel temperature,
the gap in the electronic density of states, and presumably the
ordered moments.
\end{abstract}

\normalsize


\newpage
The U-based heavy-fermion (HF) systems  URu$_2$Si$_2$,
UPd$_2$Al$_3$, and  UNi$_2$Al$_3$ exhibiting antiferromagnetic (AFM)
order followed by a superconducting transition at lower temperatures attract
much  interest in view of the possible coupling between superconducting and
magnetic order. The on the first sight similar AFM
ground state in the mentioned HF compounds reveals essential
differences. While neutron-scattering experiments
resolved an AFM ordered structure in  URu$_{2}$Si$_{2}$ with a tiny
ordered moment 0.03$\pm0.01\,\mu_{\rm B}$/U-atom below the
N\'eel temperature $T_{\rm N}$=17.5\,K
along the c-axis \cite{Brocholm87}, UPd$_{2}$Al$_{3}$
has below $T_{\rm N}$=14\,K in the basal plane  aligned
U-moments equal to 0.85$\pm0.03\,\mu_{\rm B}$ \cite{Krimmel92}.
Although the latter compound has the largest moment among the HF superconductors,
it has the largest superconducting
critical temperature $T_{\rm c}$ of about 2\,K compared to
typically 1.4\,K in URu$_{2}$Si$_{2}$. UNi$_{2}$Al$_{3}$ has been
investigated much less than the other two compounds probably  due to
specific difficulties in the preparation of good samples. This
compound is isostructural and isoelectronic to UPd$_{2}$Al$_{3}$,
but has a few times smaller magnetic moment of about
0.24$\pm0.10\,\mu_{\rm B}$ \cite{Schroder94} as well as lower
critical ($\sim 1$\,K) and N\'eel ($\sim 5$\,K) temperatures.

Pronounced anomalies in specific heat, magnetic susceptibility,
resistivity etc. for all three compounds indicate the phase
transition into an AFM state. The specific resistivity in
URu$_{2}$Si$_{2}$ has a well defined  $N$-like structure at
$T_{\rm N}$ which  looks like a kink  for UPd$_{2}$Al$_{3}$ and is
even more shallow for UNi$_{2}$Al$_{3}$. For the interpretation of
the mentioned anomalies a transition into a spin-density wave
(SDW) state has been considered \cite{Maple86} with a partial
opening of a gap at the Fermi surface
\cite{Maple86,Palstra85,Mydosh} of about 10 mV. Tunneling
experiments which can determine the gap in the electronic DOS and
its anisotropy yield for all three compounds a gap in the range 10
to 20 mV in the basal plane \cite{Aliev91,Aarts94}. However, far
infrared absorption \cite{Degiorgi97} did not resolve any gap-like
features  for UPd$_{2}$Al$_{3}$ unlike in URu$_{2}$Si$_{2}$. For
URu$_{2}$Si$_{2}$, the most investigated compound among this class
of HF systems, it is still under discussion how the large
anomalies in the transport and thermodynamic properties at $T_{\rm
N}$ can be reconciled with the tiny ordered moments. Therefore,
the understanding of the nature of the magnetic order parameter in
the AFM state of URu$_{2}$Si$_{2}$ remains a challenge. Recent
transport and neutron scattering measurements in a high magnetic
field revealed different transition fields for the AFM order or
$T_{\rm N}$ ($\sim$ 40 T, \cite{Mentink96}) and for the tiny
staggered magnetic moments ($\sim$14 T, \cite{Mason95}). This has
led to a speculation about some additional 'hidden' magnetic order
parameter in URu$_{2}$Si$_{2}$.

To clarify some aspects of the mentioned magnetic ordered state,
we have performed a comparative point-contact study on these
U-based HF compounds in strong magnetic fields. Of the three
compounds the normal state properties have been investigated
previously using point-contact spectroscopy only for URu$_2$Si$_2$
\cite{Hassel,Escudero,Thieme,Naid98}, however not in applied
magnetic fields. The d$V$/d$I(V)$ characteristics of point
contacts with URu$_2$Si$_2$ show an N-type feature related to
local contact heating above the N\'eel temperature \cite{Naid98}
and a zero-bias maximum which has been analyzed in terms of a
partial suppression of the density of states related to an
itinerant AFM ground state \cite{Hassel,Thieme,Naid98}. The
present study allows to follow these characteristics of the AFM
ground state in a magnetic field with a complete temperature
dependent study of the phenomena.

We have investigated both homocontacts between the same HF
compounds and heterocontacts between a HF compound and normal
metals like Cu or Ag. The main difference was only in the degree
of asymmetry of the d$V$/d$I(V)$ curves with respect to
bias-voltage polarity, which is more pronounced for
heterocontacts. The origin of the asymmetry is still under
discussion \cite{Naid98}. Because  this effect has no influence
on the main conclusions of the present investigations, we will
put no more attention to it. In the case of the URu$_{2}$Si$_{2}$
single crystal, the heterocontacts were established in such a way
that both contact axis and magnetic field were parallel to the
c-axis or perpendicular to it. For the UPd$_{2}$Al$_{3}$ single
crystal the contact axis and magnetic field were aligned along
the easy basal plane direction whereas we used UNi$_{2}$Al$_{3}$
samples of unknown orientation. The measurements were carried in
magnetic fields up to 28~T at 4.2\,K, but for UNi$_{2}$Al$_{3}$
down to about 2\,K and up to 10\,T.

\begin{figure}
\includegraphics [width=12cm]{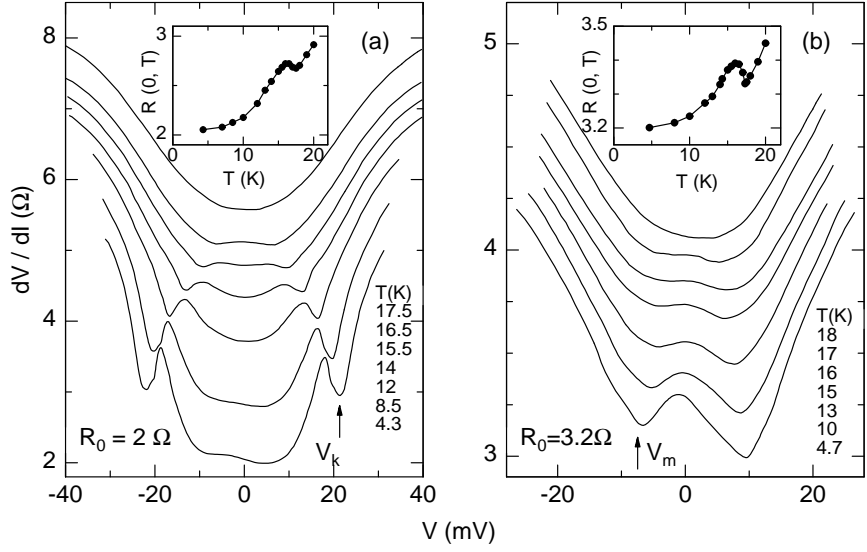}
\caption{Two types of behaviour, (a) and (b), in the d$V$/d$I(V)$
curves for single crystal URu$_{2}$Si$_{2}$ homocontacts
established in the basal plane at increasing temperature up to
$T_{\rm N}$. The curves are offset vertically for clarity. The
insets show the temperature dependence of the zero bias
resistance, which mimics $\rho (T)$ for bulk samples. }
\label{urs2type}
\end{figure}

The measured d$V$/d$I(V)$ curves of URu$_{2}$Si$_{2}$  contacts
can be separated into three groups. In the first group the
d$V$/d$I(V)$ curves mimic the behaviour of bulk $\rho(T)$. The
differential resistance increases with voltage and exhibits a
$N$-type feature  at about 20 mV (Fig.\,\ref{urs2type}a) similar
to $\rho(T)$ at $T_{\rm N}$ \cite{Mydosh}. The second type of
d$V$/d$I(V)$ reveals a pronounced asymmetric zero-bias maximum
(ZBM) of a width of about 10 mV  in d$V$/d$I(V)$
(Fig.\,\ref{urs2type}b) followed by  gradually increasing signal
at higher voltages. The third one contains simultaneously
both kinds of structures in d$V$/d$I(V)$. We note
that the temperature dependence of the contact resistance
(Fig.\,\ref{urs2type}, insets) corresponds in all cases to $\rho
(T)$ independent of the type of d$V$/d$I(V)$ behaviour. This
indicates that the material in the constriction reflects the bulk
properties. The mentioned features, namely $N$-type kink and ZBM,
vanish at the N\'eel temperature (Fig.\,\ref{urs2type}) and
are therefore connected with the magnetic state.

\begin{figure}
\includegraphics [width=12cm]{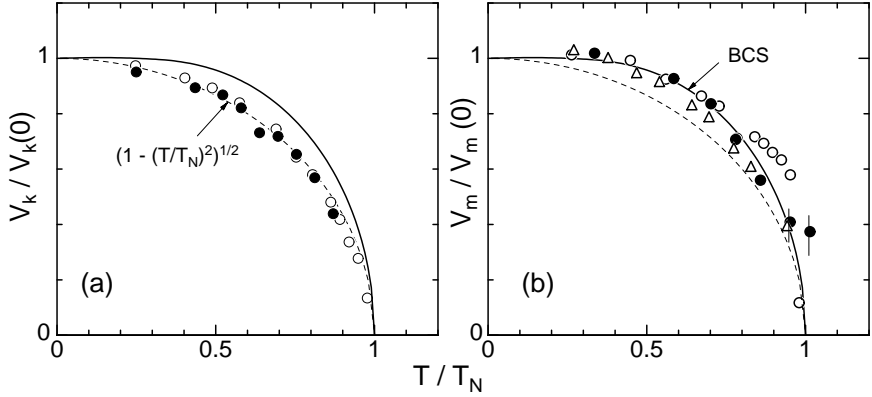}
\caption{Temperature dependence of the reduced voltage positions
$V_{\rm k}$ (a) and  $V_{\rm m}$ (b) (see for definition
Fig.\,\ref{urs2type}) for a few URu$_{2}$Si$_{2}$ homocontacts
established in the basal plane. The solid circles for both figures
correspond to the same homocontact. The solid line in both figures
is the mean-field BCS dependence while the dashed curve describes
the thermal regime behaviour \cite{Naid98}. } \label{urs2del}
\end{figure}

The voltage position of the $N$-kink (marked by $V_{\rm k}$ in
Fig.\,\ref{urs2type}a) is determined by $T_{\rm N}$ and
corresponds to the transition of the contact region from the AFM
to the paramagnetic state most likely due to bias-voltage heating
in the constriction. The temperature dependence $\sqrt{1-(T/T_{\rm
N})^2}$ of $V_{\rm k}$ shown in  Fig.\,\ref{urs2del}a is expected
for such a local contact heating \cite{Naid98}. The ZBM is more
pronounced for curves with shallow or not resolved kink
peculiarities. Moreover, the ZBM cannot be described in the
thermal model what can be directly seen upon comparing
d$V$/d$I(V>0)$ with d$V$/d$I(V=0,T)=R(V=0,T)$ (see Fig.1b). These
observations point to the spectral nature of ZBM. The latter has
been related (see e. g. \cite{Naid98}) with the existence of a gap
in the excitation spectrum of the electrons due to the formation
of a SDW below $T_{\rm N}$. The ZBM has a width which is
comparable with the gap value estimated in
\cite{Maple86,Palstra85,Mydosh,Aarts94}. The intensity of the ZBM
gradually decreases with increasing temperature
\cite{Hassel,Naid98} analogously to the intensity of AFM Bragg
peaks describing the behaviour of the staggered magnetic moments
or the order parameter \cite{Dijk97}. Therefore, it is tempting to
connect the ZBM also with the magnetic order parameter, although
the microscopic nature of the tiny staggered magnetic moments in
URu$_2$Si$_2$  as well as its influence on the measured d$V/$d$I$
are still unknown. Because the intensity  of  ZBM depends on the
chosen criterion for the subtraction of the increased with a
voltage background, we suggest to take  voltage position of the
minima $V_{\rm m}$ (see Fig.\,\ref{urs2type}b) as an additional
measure for the magnetic order parameter as supported by the
mean-field (BCS-like) $V_{\rm m}(T)$ dependence from
Fig.\,\ref{urs2del}b as also found in \cite{Escudero,Thieme}.

\begin{figure}
\includegraphics [width=12cm]{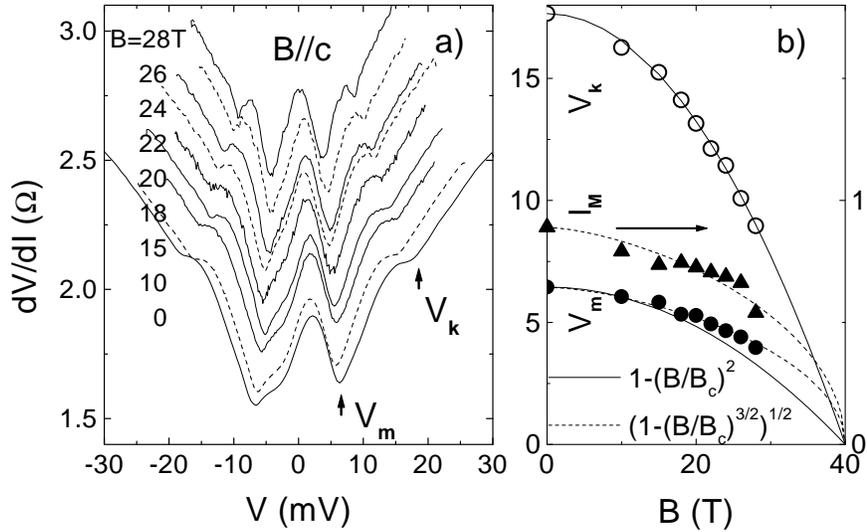}
\caption{(a) d$V$/d$I\,(V)$ curves for a URu$_2$Si$_2$-Cu
heterocontact in magnetic fields along the easy $c$ axis  at
$T$=4.2\,K. The solid curves correspond to the field sweep up,
while the dashed one to the field sweep down. The arrows show
definition of the kink $V_{\rm k}$ and minimum $V_{\rm m}$
positions. The curves are offset vertically for clarity. (b)
Dependence of $V_{\rm k}$, $V_{\rm m}$ (left scale) and ZBM
integrated intensity $I_{\rm M}$ (right scale) versus magnetic
field. Note, position of $V_{\rm k}$, $V_{\rm m}$ and ZBM
intensity was taken after   symmetrizing of  d$V$/d$I\,(V)$
curves.  The solid lines represent dependence $(1-(B/B_{\rm
c})^2)$ characteristic for $T_{\rm N}(B)$ and spin-wave gap
$\Delta(B)$ \cite{Mentink96} while dashed line $\sqrt{1-(B/B_{\rm
c})^{3/2}}$ is taken from \cite{Mason95} for staggered magnetic
moments.} \label{ursh1}
\end{figure}

The point-contact data presented in Fig.\,\ref{ursh1}a for
magnetic fields parallel to the easy c-axis exhibit both types of
features discussed above, i. e. ZBM and N-kink. The integrated
intensity of the ZBM (after subtracting a polynomal
voltage-dependent background) is close to the $\sqrt{1-(B/B_{\rm
c})^{3/2}}$ behaviour like that for magnetic moments
\cite{Mason95}. As shown in Fig.\,\ref{ursh1}b, the same
dependence is found for $V_{\rm m}$ as well. On the other hand,
$V_{\rm k}$ follows the magnetic field dependence $(1-(B/B_{\rm
c})^2)$ like for $T_{\rm N}$ \cite{Mentink96}(shown in
Fig.\,\ref{ursh1}b). The latter dependence is found also for the
width of ZBM (not shown), which is related to the SDW gap. Thus,
the mentioned features in d$V$/d$I(V)$ - $V_{\rm k}$, ZBM width
and ZBM intensity or $V_{\rm m}$ measured on the same contact -
are described by the magnetic field dependencies characteristic
for the behaviour of transition temperature $T_{\rm N}$, magnetic
gap width \cite{Mentink96} and magnetic order parameter
\cite{Mason95}, respectively. Moreover, as can be seen in
Fig.\,\ref{ursh1}b, independent of the type of behaviour in all
cases the critical field is estimated to be about 40~T what
coincides with $B_{\rm c}$ values measured by magnetoresistance
\cite{Sugiyama}. Therefore, unlike in \cite{Mentink96,Mason95}
where for the ordered magnetic moments the critical field is
estimated to be about 14\,T our data show the presence of one
order parameter, which vanishes at $T_{\rm N}$=17.5 K and $B_{\rm
c}\simeq$40 T. This is in line with the recent observation of van
Dijk et al. \cite{Dijk97} that the ordered moments remain coupled
to the energy gap in the magnetic excitation spectrum in fields at
least up to 12 T. We should emphasize that by  measuring
URu$_2$Si$_2$ contacts in a field perpendicular to the easy c-axis
direction we did not found any remarkable influence of a magnetic
field on d$V$/d$I(V)$ testifying that the point contact data
really reflect the bulk properties.

\begin{figure}
\includegraphics [width=10cm]{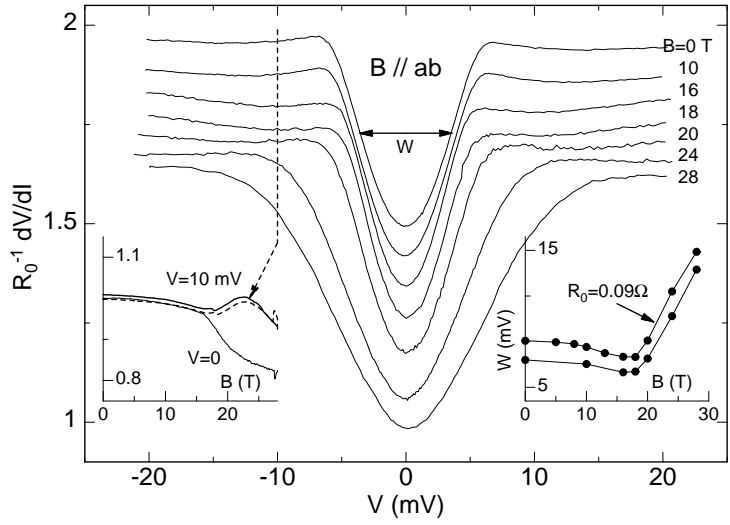}
\caption{ d$V$/d$I(V)$ curves for a UPd$_{2}$Al$_{3}$-Cu
heterocontact with $R_0$=4.3 $\Omega$ at different magnetic fields
along the basal plane and $T$=4.2\,K. Horizontal line with arrows
shows determination of the width of the minimum. The curves are
offset vertically for clarity. Right inset: width of the minimum
versus magnetic filed for the previous contact  and for another
contact with $R_0$=0.09 $\Omega$. Left inset: magnetoresistance of
the contact  with $R_0$=4.3 $\Omega$ at zero bias and at 10\,mV.
} \label{upa}
\end{figure}

Let us turn to the other compounds. The d$V$/d$I(V)$ curve of
UPd$_{2}$Al$_{3}$ contacts (see Fig.\,\ref{upa}) show a minimum at $V$=0
with edge maxima or shoulders which are connected with the AFM transition due
to the heating effect \cite{Kvit99}, analogously to the $N$-type feature
in the case of URu$_{2}$Si$_{2}$. However, for UPd$_{2}$Al$_{3}$
contacts we have never seen even a shallow ZBM neither for
homo- nor for heterocontacts after the study of more than 20 contacts
both below and above the N\'eel temperature. Therefore no evidence of
the partially gapping of the Fermi surface is observed for
UPd$_{2}$Al$_{3}$ unlike in URu$_{2}$Si$_{2}$ what points to a quite different
magnetic ground state as well as to the different nature of the ordered
moments for both compounds.

A magnetic field along the easy basal plane modifies the
d$V$/d$I(V)$ curves of UPd$_{2}$Al$_{3}$ as can be seen from
Fig.\,\ref{upa}a. The maxima slightly shift ($\approx 15~\%$) to
lower energies and broaden with increasing magnetic field  up to
18\,T, and than vanish in higher fields. The width of the
d$V$/d$I(V)$ minimum at $V$=0 has a minimum at 18\,T, while the
contact resistance has a kink at this field both at zero bias and
finite bias voltage (see Fig.\,\ref{upa}b,c). Hence the
metamagnetic transition at 18\,T \cite{deViss93} is clearly
resolved in point contact measurements, while  no other phase
boundary was observed both at lower and higher fields up to
28\,T.  From measurements of the dc susceptibility, dc
magnetization, transverse magnetoresistivity,  and
magnetostriction,  Grauel et al. \cite{Grauel92} have also found
a phase boundary in UPd$_{2}$Al$_{3}$ at a critical field of
about 4 T along the base plane.  However, the  influence of this
low-field transition on the resistivity  is at least one order of
magnitude smaller compared to the transition at 18\, T. Moreover,
de Visser  et al. \cite{deViss93} did not found a 4-T  transition
in their magnetoresistance data indicating that a re-orientation
of the AFM domains could play a role in this phenomenon.

\begin{figure}
\includegraphics [width=9cm]{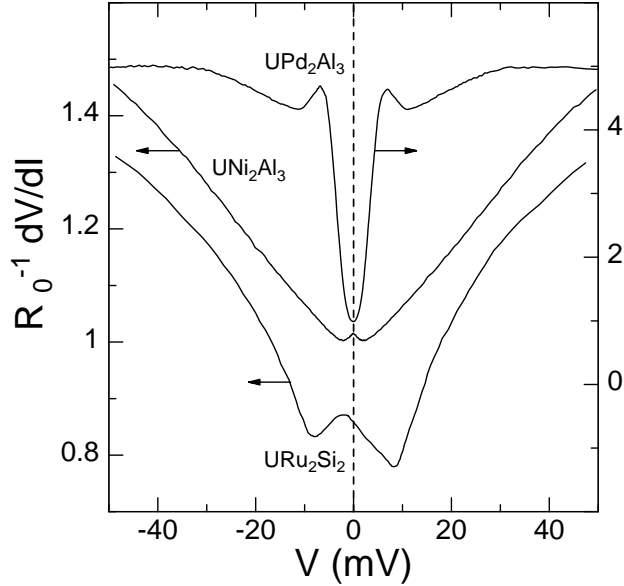}
\caption{Comparison of d$V$/d$I(V)$ curves for homocontacts with
three studied HF compounds: UPd$_{2}$Al$_{3}$ ($R_0$=0.61
$\Omega$, $T$=4.2\,K), UNi$_{2}$Al$_{3}$ ($R_0$=1.5 $\Omega$, $T$=
2.3\,K) and URu$_{2}$Si$_{2}$ ($R_0=3.2 \Omega$, $T$=4.2\,K). A
ZBM is only resolved for the two latter compounds. The curve for
URu$_{2}$Si$_{2}$ is shifted down by 0.15. } \label{hfs}
\end{figure}

The  d$V$/d$I(V)$ curves of UNi$_{2}$Al$_{3}$ represent usually a
smooth broad almost symmetric minimum around zero-bias. However
often a shallow ZBM can be observed around $V$=0
(Fig.\,\ref{hfs}). The distance between the  minima  in
d$V$/d$I(V)$ with ZBM is about a few mV (often up to 10 mV) and
ZBM disappears at about 5\,K (between 10-15\,K for wider maxima).
For ZBM with critical temperature of about 5\,K critical field was
about 10\,T. From this point of view UNi$_{2}$Al$_{3}$ behaves
similar to URu$_{2}$Si$_{2}$ what hints to the developing of a
magnetic state with partially gapping of the Fermi surface in this
compound too. It should be noted that we didn't resolve any
feature in d$V$/d$I(V)$ for UNi$_2$Al$_3$ (Fig.\,\ref{hfs})
connected with  $T_{\rm N}$ like that in URu$_2$Si$_2$
(Fig.\,\ref{urs2type}a) and UPd$_2$Al$_3$ (Fig.\,\ref{hfs}). This
transition at $T_{\rm N}$ is also very shallow in the $\rho (T)$
dependence of UNi$_2$Al$_3$. Probably, a better quality of the
UNi$_2$Al$_3$ samples is required to register the AFM transition
and to study  the temperature behaviour of ZBM in d$V$/d$I(V)$.

Summarizing, the point-contact measurements for the investigated
U-based heavy fermion compounds yield information on the
differences in the AFM ground state of these systems. The
ZBM-structure  in d$V$/d$I(V)$ for the URu$_{2}$Si$_{2}$ contacts
points to a partially gapped Fermi surface in the magnetically
ordered state, but no evidence of an analogous structure has been
found in the case of UPd$_2$Al$_3$ unlike for UNi$_2$Al$_3$ where
it is possible to resolve a shallow ZBM. The results for
URu$_{2}$Si$_{2}$ in the $H-V,T$ diagram yield only one critical
N\'eel temperature of 17\,K and one critical field of about 40\,T
along the easy c-axis for all features in d$V$/d$I(V)$ testifying
that they result from the same order parameter in the magnetic
state.

\end{document}